\documentclass[reprint,nofootinbib]{revtex4}
\usepackage{graphicx}  
\usepackage{dcolumn}   
\usepackage{bm}        
\usepackage{amssymb}
\usepackage{hyperref}
\usepackage{amsmath}

\begin{document}

\title{\bf Model independent analysis of New Physics effects on $B_c \to (D_s,\,D^{\ast}_s)\,\mu^+\mu^-$ decay observables}
\author{Rupak~Dutta${}^{}$}
\email{rupak@phy.nits.ac.in}
\affiliation{
${}$National Institute of Technology Silchar, Silchar 788010, India\\
}

\begin{abstract}
Motivated by the anomalies present in $b \to s\,l^+\,l^-$ neutral current decays, we study the corresponding 
$B_c \to (D_s,\,D^{\ast}_s)\mu^+\mu^-$ decays within the standard model and beyond. We use a model independent effective theory formalism 
in the presence
of vector and axial vector new physics operators and study the implications of the latest global fit to the $b \to s\,l^+\,l^-$ data on
various observables for the $B_c \to (D_s,\,D^{\ast}_s)\mu^+\mu^-$ decays. We give predictions on several observables such as
the differential branching ratio, ratio of branching ratios, forward backward asymmetry, and the longitudinal polarization fraction of the 
$D^{\ast}_s$ meson within standard model and within various new physics scenarios. These results can be tested at the Large Hadron Collider
and, in principle, can provide complimentary information regarding new physics in $b \to s\,l^+\,l^-$ neutral current decays.

\end{abstract}

\maketitle

\section{Introduction}
\label{int}
Although standard model~(SM) of particle physics is successful in explaining various experimental observations, it, however, can not
accommodate several long standing issues such as dark matter, dark energy, neutrino mass, matter antimatter asymmetry in the universe etc. 
It is indeed certain that physics beyond the
SM exists. There are two ways to determine the nature of new physics~(NP). One is direct detection of new particles and their interactions and
another is indirect detection through their effects on various low energy processes. In this respect, flavor physics can, in principle, be 
the ideal platform to look for indirect evidences of NP. In fact, various anomalies with the SM prediction have been reported by dedicated 
experiments such as BABAR, Belle, and more recently by LHCb. In particular, measurement of various observables in $b \to c\tau\nu$ charged
current interactions and in $b \to s\,l^+\,l^-$ neutral current interactions already provided hints of NP. We will focus here on anomalies 
present in $B$ meson decays mediated via $b \to s\,l^+\,l^-$ neutral current interactions. The most important observables are the lepton
flavor universality~(LFU) ratios $R_K$ and $R_{K^{\ast}}$, various angular observables in $B \to K^{\ast}\mu^+\mu^-$ decays, and the branching
ratio of $B_s \to \phi\mu^+\mu^-$ decays. The experimental results confirming these anomalies
are listed below.

A significant deviation from the SM expectation is observed in the LFU ratios $R_K$ and $R_{K^{\ast}}$ defined as
\begin{eqnarray}
R_K^{(*)} = \frac{\mathcal B(B \to K^{(*)}\,\mu\mu)}{\mathcal B(B \to K^{(*)}\,e\,e)}\,.
\end{eqnarray}
The first LHCb measurement of $R_K=0.745^{+0.090}_{-0.074}\pm0.036$~\cite{Aaij:2014ora} in the low $q^2$ bin $1< q^2 < 6\,{\rm GeV^2}$ 
deviates from the SM prediction $R_K \approx 1$~\cite{Hiller:2003js,Bobeth:2007dw,Bordone:2016gaq} at $2.6\sigma$ 
level. Very recently, the earlier measurement was superseded by LHCb Collaboration and it is reported to be 
$R_K=0.846^{+0.060+0.016}_{-0.054-0.014}$~\cite{Aaij:2019wad}.
Although it moves closer to the SM value, the deviation with the SM prediction still stands at $2.5\sigma$ level.
Similarly, the measured value of
$R_{K^{\ast}} = 0.66^{+0.11}_{-0.07}\pm 0.03$ and $0.69^{+0.11}_{-0.07}\pm0.05$ in the dilepton invariant mass $q^2 = [0.045,\,1.1]\,
{\rm GeV^2}$ and $[1.1,\,6.0]\,{\rm GeV^2}$~\cite{Aaij:2017vbb} 
deviate from the SM prediction of $R_{K^{\ast}} \approx 1$~\cite{Capdevila:2016ivx,Serra:2016ivr} at approximately $2.1\sigma$ and 
$2.4\sigma$, respectively. Very recently, Belle collaboration has reported
the values of $R_{K^{\ast}}$ in multiple $q^2$ bin but with a much larger uncertainties~\cite{Abdesselam:2019wac}. The other notable 
deviation is the deviation
observed in the angular observable $P^{\prime}_5$ in $B \to K^{\ast}\mu^+\mu^-$ decays~\cite{DescotesGenon:2012zf,Descotes-Genon:2013vna}. 
LHCb~\cite{Aaij:2013qta,Aaij:2015oid} and ATLAS~\cite{Aaboud:2018krd} have measured the value
of the angular observable $P^{\prime}_5$ in the $q^2$ range $4.0<q^2<6.0\,{\rm GeV^2}$ and 
the deviation from the SM prediction is found to be more than $3\sigma$~\cite{Aebischer:2018iyb}. Belle~\cite{Abdesselam:2016llu} and 
CMS~\cite{cms} have also measured this observable in the
$q^2$ bin $4.3<q^2<8.68\,{\rm GeV^2}$ and $4.3<q^2<6.0\,{\rm GeV^2}$, respectively. Although the Belle measured value differs from the SM
expectation at $2.6\sigma$ level, the measured value by CMS is consistent with the SM expectation at $1\sigma$ level. Similarly, there is
a systematic deficit in the measured value of branching ratio of $B_s \to \phi\mu^+\mu^-$~\cite{Aaij:2013aln,Aaij:2015esa} decays as 
compared to the SM prediction~\cite{Aebischer:2018iyb,Straub:2015ica}. Currently
the deviation with the SM prediction stands at around $3.7\sigma$. If it persists and is confirmed by future 
experiments, it could unravel new flavor structure beyond the SM physics. Various global fits~\cite{Capdevila:2017bsm,Altmannshofer:2017yso,
DAmico:2017mtc,Hiller:2017bzc,Geng:2017svp,Ciuchini:2017mik,Celis:2017doq,Alok:2017sui,Alok:2017jgr,Alok:2019ufo} to the $b \to s\,l^+\,l^-$
data have been performed and it was suggested that some of these anomalies can be resolved by modifying the Wilson coefficients~(WCs).  

If these anomalies are due to NP, this will show up in other $b \to s\,l^+\,l^-$ transition decays as well. In this paper, we analyze 
$B_c \to (D_s,\,D^{\ast}_s)\mu^+\mu^-$ decays mediated via $b \to s\,l^+\,l^-$ neutral current transitions within the SM and in several
NP scenarios. LHCb has already measured the ratio of branching ratio $R_{J/\Psi}$
in $B_c \to J/\Psi\,l\,\nu$ decays. Detection and measurement of various observables pertaining to $B_c$ meson decaying to other mesons
via $b \to s\,l^+\,l^-$ neutral current interactions will be feasible once more and more data will be accumulated by LHCb. It is worth 
mentioning that the study of such modes is complimentary to the study of $B \to (K,\,K^{\ast})\mu^+\mu^-$ decays and it can, in principle, 
provide useful information regarding different NP Lorentz structures.
Moreover, study of these decay modes both theoretically and experimentally can act as a useful ingredient in maximizing future sensitivity to 
NP.   

Within the SM, $B_c \to (D_s,\,D^{\ast}_s)\mu^+\mu^-$ decays have been studied previously using the relativistic constituent quark 
model~\cite{Faessler:2002ut},  light-front quark model~\cite{Geng:2001vy,Choi:2010ha}, QCD sum rules~\cite{Azizi:2008vy,Azizi:2008vv}, and 
relativistic quark model~\cite{Ebert:2010dv}. In this paper, we use
the relativistic quark model of Ref.~\cite{Ebert:2010dv} and supplement the previous analysis 
by analyzing the effect of various NP on these decay modes in a model independent way. We use an effective theory formalism
in the presence of new vector and axial vector couplings that couples only to the muon sector. We give prediction of various observables
such as the ratio of branching ratios, lepton side forward backward asymmetry, and the longitudinal polarization fraction of the $D^{\ast}_s$
meson within the SM and within various NP scenarios.

Our paper is organized as follows. In section~\ref{form}, we start with the effective weak Hamiltonian for $b \to s\,l^+\,l^-$ decays
in the presence of new vector and axial vector operators. We also discuss the hadronic matrix elements of $B_c \to D_s$ and 
$B_c \to D^{\ast}_s$ and their parametrization in terms of various meson to meson transition form factors. In section~\ref{hel}, we write
down the helicity amplitudes for the $B_c \to D_s\mu^+\mu^-$ and $B_c \to D^{\ast}_s\mu^+\mu^-$ decay modes and construct several observables.
In section.~\ref{res}, we give predictions of all the observables in the SM and in several NP cases obtained from the global fit. We 
conclude with a brief summary of our results in section.~\ref{con}.
\section{Formalism}
\label{form}
The most general effective weak Hamiltonian in the presence of new vector and axial vector operators for the $|\Delta B|=|\Delta S|=1$ 
transition can be written as
\begin{eqnarray}
\mathcal H_{\rm eff} &=& -\frac{4\,G_F}{\sqrt{2}}\,V_{tb}\,V_{ts}^{\ast}\,\frac{\alpha_e}{4\,\pi}\Big[C_9^{\rm eff}\bar{s}\gamma^{\mu}\,
P_L\,b\,
\bar{l}\gamma_{\mu}\,l+C_{10}\,\bar{s}\gamma^{\mu}\,P_L\,b\,\bar{l}\gamma_{\mu}\,\gamma_5\,l - \frac{2\,m_b}{q^2}\,C_7^{\rm eff}\,\bar{s}i\,
q_{\nu}\,\sigma^{\mu\nu}\,P_R\,b\,\bar{l}\gamma_{\mu}\,l \nonumber \\
&&+ C_9^{NP}\,\bar{s}\gamma^{\mu}\,P_L\,b\,\bar{l}\gamma_{\mu}\,l+C_{10}^{NP}\,
\bar{s}\gamma^{\mu}\,P_L\,b\,\bar{l}\gamma_{\mu}\,\gamma_5\,l + C^{\prime}_9\,\bar{s}\gamma^{\mu}\,P_R\,b\,\bar{l}\gamma_{\mu}\,l +
C^{\prime}_{10}\,\bar{s}\gamma^{\mu}\,P_R\,b\,\bar{l}\gamma_{\mu}\,\gamma_5\,l\Big]\,,
\end{eqnarray}
where $G_F$ is the Fermi coupling constant, $\alpha_e$ is the electromagnetic coupling constant, $V_{tb}$ and $V_{ts}$ are the relevant 
Cabibbo~Kobayashi~Maskawa~(CKM)
matrix elements, and $P_{R,\,L} = (1\pm \gamma_5)/2$ are the chiral projectors. All the WCs are evaluated at a 
renormalization scale of $\mu = m^{\rm pole}_b = 4.8\,{\rm GeV}$. The $b$ quark mass associated with $C_7^{\rm eff}$ is considered to be 
running
mass in the $\overline{\rm MS}$ scheme. In principle, there can be several
NP Lorentz structures such as vector, axial vector, scalar, pseudoscalar, and tensor. The scalar, pseudoscalar and the tensor NP operators
are severely constrained by $B_s \to \mu\mu$ and $b \to s\gamma$ measurements~\cite{Alok:2010zd,Alok:2011gv,Bardhan:2017xcc}. Hence, we 
consider NP in the form of
vector and axial vector operators only. Again, we do not consider NP in the dipole operator as these are well constrained by radiative
decays. The non factorizable corrections coming from electromagnetic corrections to the matrix elements of purely hadronic
operators in the weak effective Hamiltonian are ignored in our analysis. These corrections, however, are expected to be significant at low
$q^2$~\cite{Beneke:2001at,Beneke:2004dp}. All the NP WCs $C_9^{NP}$, $C_{10}^{NP}$, $C^{\prime}_9$, and $C^{\prime}_{10}$ 
are assumed to be real for our
analysis. In the SM, $C_9^{NP}=C_{10}^{NP}=C^{\prime}_9=C^{\prime}_{10}=0$. The effective WCs
$C_7^{\rm eff}$ and $C_9^{\rm eff}$ are defined as
\begin{eqnarray}
&&C_7^{\rm eff} = C_7 - \frac{1}{3}\,C_5 - C_6\,, \nonumber \\
&&C_9^{\rm eff} = C_9 + y(q^2) + y_{\rm BW}(q^2)\,,
\end{eqnarray}
where the contributions coming from the one loop matrix elements of the four quark operators are contained in~\cite{Buras:1994dj} 
\begin{eqnarray}
y(q^2)&=&h\Big(\frac{m_c}{m_b},\,\frac{q^2}{m_b^2}\Big)(3\,C_1+C_2+3\,C_3+C_4+3\,C_5+C_6) -\frac{1}{2}\,h\Big(1,\,\frac{q^2}{m_b^2}\Big)
(4\,C_3+4\,C_4+3\,C_5+C_6) \nonumber \\
&&- 
\frac{1}{2}\,h\Big(0,\,\frac{q^2}{m_b^2}\Big)(C_3+3\,C_4)+\frac{2}{9}(3\,C_3+C_4+3\,C_5+C_6)\,.
\end{eqnarray}
Here
\begin{eqnarray}
&&h(z,\,s)=-\ln\frac{m_b}{\mu}-\frac{8}{9}\ln\,z+\frac{8}{27}+\frac{4}{9}x-\frac{2}{9}(2+x)|1-x|^{1/2}
\begin{cases}
\ln\Big|\frac{\sqrt{1-x}+1}{\sqrt{1-x}-1}\Big|-i\,\pi & \rm{for}\,\, x=\frac{4\,z^2}{s}<1\\
2\,\rm{arctan}\,\frac{1}{\sqrt{x-1}} & \rm{for}\,\, x=\frac{4\,z^2}{s}>1
\end{cases}
\nonumber \\
&&h(0,\,s)=\frac{8}{27}-\ln\frac{m_b}{\mu}-\frac{4}{9}\ln\,s+\frac{4}{9}\,i\,\pi\,.
\end{eqnarray}
The phenomenological parameter $y_{\rm BW}(q^2)$ involves the long distance effects coming from the $c\bar{c}$ resonance contributions
coming from $J/\Psi$, $\Psi^{\prime}$ etc. In particular, these resonances provide large peaked contributions in the $q^2$ bins that are
close to these charmonium resonance masses. The corresponding $q^2$ bins are not considered in our analysis.
The values of masses of charm and bottom quark in these expressions are defined in pole mass scheme. The WCs
that contains the short distance contribution can be calculated perturbatively, whereas, for the calculation of the long distance 
contributions contained in the matrix elements of local operators between initial and final hadron states, it requires non perturbative
approach. The hadronic matrix elements can be expressed in terms of various meson to meson transition form factors.

The hadronic matrix elements for the $B_c \to D_s\mu^+\mu^-$ decays can be parametrized in terms of three invariant form factors. Those are 
\begin{eqnarray}
<D_s|\bar{s}\gamma^{\mu}b|B_c> &=& f_{+}(q^2)\Big[p_{B_c}^{\mu}+p_{D_s}^{\mu}-\frac{M_{B_c}^2-M_{D_s}^2}{q^2}\,q^{\mu}\Big] + f_0(q^2)
\frac{M_{B_c}^2-M_{D_s}^2}{q^2}\,q^{\mu}\,,\nonumber \\
<D_s|\bar{s}\sigma^{\mu\nu}\,q_{\nu}b|B_c> &=&\frac{i\,f_T(q^2)}{M_{B_c}+M_{D_s}}\Big[q^2(p_{B_c}^{\mu}+p_{D_s}^{\mu} - (M_{B_c}^2-
M_{D_s}^2)q^{\mu}\Big]\,.
\end{eqnarray}
Similarly, for the $B_c \to D^{\ast}_s\mu^+\mu^-$ decays, the hadronic matrix elements can be parametrized in terms of seven invariant
form factors, i.e,
\begin{eqnarray}
<D^{\ast}_s|\bar{s}\gamma^{\mu}b|B_c> &=& \frac{2\,i\,V(q^2)}{M_{B_c}+M_{D^{\ast}_s}}\,\epsilon^{\mu\nu\rho\sigma}\epsilon^{\ast}_{\nu}\,
p_{B_{c_{\rho}}}\,p_{{D^{\ast}_s}_{\sigma}}\,, \nonumber \\
<D^{\ast}_s|\bar{s}\gamma^{\mu}\gamma_5\,b|B_c> &=& 2\,M_{D^{\ast}_s}\,A_0(q^2)\frac{\epsilon^{\ast}\cdot q}{q^2}\,q^{\mu} + 
(M_{B_c} + M_{D^{\ast}_s})\,A_1(q^2)\Big(\epsilon^{{\ast}^{\mu}}-\frac{\epsilon^{\ast}\cdot q}{q^2}\,q^{\mu}\Big)\, \nonumber \\
&&-
A_2(q^2)\frac{\epsilon^{\ast}\cdot q}{M_{B_c} + M_{D^{\ast}_s}}\,\Big[p_{B_c}^{\mu}+p_{D^{\ast}_s}^{\mu}-\frac{M_{B_c}^2-M_{D^{\ast}_s}^2}
{q^2}\,q^{\mu}\Big]\,, \nonumber \\
<D^{\ast}_s|\bar{s}\,i\,\sigma^{\mu\nu}\,q_{\nu}b|B_c> &=& 2\,T_1(q^2)\,\epsilon^{\mu\nu\rho\sigma}\epsilon^{\ast}_{\nu}\,p_{B_{c_{\rho}}}\,
p_{{D^{\ast}_s}_{\sigma}}\,, \nonumber \\
<D^{\ast}_s|\bar{s}\,i\,\sigma^{\mu\nu}\,\gamma_5\,q_{\nu}b|B_c> &=& T_2(q^2)\,\Big[(M_{B_c}^2-M_{D^{\ast}_s}^2)\epsilon^{{\ast}^{\mu}} - 
(\epsilon^{\ast}\cdot q)(p_{B_c}^{\mu}+p_{D^{\ast}_s}^{\mu})\Big] \nonumber \\
&&+ 
T_3(q^2)\,(\epsilon^{\ast}\cdot q)\Big[q^{\mu}-\frac{q^2}
{M_{B_c}^2-M_{D^{\ast}_s}^2}(p_{B_c}^{\mu}+p_{D^{\ast}_s}^{\mu})\Big]\,,
\end{eqnarray}
where $q^{\mu}=(p_B-p_{D_s,\,D^{\ast}_s})^{\mu}$ is the four momentum transfer and $\epsilon_{\mu}$ is polarization vector of the 
$D^{\ast}_s$ meson. For the $B_c \to D_s$ and $B_c \to D^{\ast}_s$ transition form factors we follow the relativistic quark model adopted in
Ref.~\cite{Ebert:2010dv}. It was mentioned in Ref.~\cite{Ebert:2010dv} that in the limit of infinitely heavy quark mass and large
energy of the final meson, the form factor results obtained in this approach are consistent with all the model independent symmetry
relations~\cite{Charles:1998dr,Ebert:2001pc}. We refer to Ref.~\cite{Ebert:2010dv} for all the omitted details.
\section{Helicity amplitudes and decay observables}
\label{hel}
For the helicity amplitudes, we pattern our analysis after that of Ref.~\cite{Ebert:2010dv} and, indeed, adopt a common notation.
We use the helicity techniques of Refs.~\cite{Korner:1989qb,Kadeer:2005aq} and write the hadronic helicity amplitudes for 
$B_c \to D_s\,l^+\,l^-$ decays in the presence of vector and axial vector NP operators as follows:
\begin{eqnarray}
\label{dseq}
&&H_{\pm}^{(i)}=0 \,, \nonumber \\
&&H_0^{(1)}=\sqrt{\frac{\lambda}{q2}}\Big[(C_9^{\rm eff}+C_9^{NP}+C^{\prime}_9)\,f_{+}(q^2) + C_7^{\rm eff}\,\frac{2\,m_b}
{M_{B_c}+M_{D_s}}\,f_T(q^2)\Big]\, \nonumber \\
&&H_0^{(2)}=\sqrt{\frac{\lambda}{q2}}\,(C_{10}+C_{10}^{NP}+C^{\prime}_{10})\,f_{+}(q^2)\,, \nonumber \\
&&H_t^{(1)}=\frac{M_{B_c}^2-M_{D_s}^2}{q^2}\,(C_9^{\rm eff}+C_9^{NP}+C^{\prime}_9)\,f_0(q^2)\,, \nonumber \\
&&H_t^{(2)}=\frac{M_{B_c}^2-M_{D_s}^2}{q^2}\,(C_{10}+C_{10}^{NP}+C^{\prime}_{10})\,f_0(q^2)\,
\end{eqnarray}
Similarly, for $B_c \to D^{\ast}_s\,l^+\,l^-$ decays, the hadronic helicity amplitudes are
\begin{eqnarray}
H_{\pm}^{(1)}&=&-(M_{B_c}^2-M_{D^{\ast}_s}^2)\Big[(C_9^{\rm eff}+C_9^{NP}-C^{\prime}_9)\,\frac{A_1(q^2)}{M_{B_c} - M_{D^{\ast}_s}} + 
\frac{2\,m_b}{q^2}\,C_7^{\rm eff}\,T_2(q^2)\Big] \nonumber \\
&&\pm 
\sqrt{\lambda}\Big[(C_9^{\rm eff}+C_9^{NP}+C^{\prime}_9)\,\frac{V(q^2)}{M_{B_c}+M_{D^{\ast}_s}} + \frac{2\,m_b}
{q^2}\,C_7^{\rm eff}\,T_2(q^2)\Big]\,, \nonumber \\
H_{\pm}^{(2)}&=&(C_{10}+C_{10}^{NP}-C^{\prime}_{10})\Big[-(M_{B_c}+M_{D^{\ast}_s})\,A_1(q^2)\Big] \pm (C_{10}+C_{10}^{NP}+C^{\prime}_{10})
\frac{\lambda}{M_{B_c}+M_{D^{\ast}_s}}\,V(q^2)\,, \nonumber \\
H_0^{(1)}&=& -\frac{1}{2\,M_{D^{\ast}_s}\,\sqrt{q^2}}\Bigg\{(C_9^{\rm eff}+C_9^{NP}-C^{\prime}_9)\,\Bigg[(M_{B_c}^2-M_{D^{\ast}_s}^2-q^2)\,
(M_{B_c}+M_{D^{\ast}_s})A_1(q^2) - \frac{\lambda}{M_{B_c}+M_{D^{\ast}_s}}\,A_2(q^2)\Bigg]\, \nonumber \\
&&+
2\,m_b\,C_7^{\rm eff}\,\Bigg[(M_{B_c}^2+3\,M_{D^{\ast}_s}^2-q^2)\,T_2(q^2) - \frac{\lambda}{M_{B_c}^2-M_{D^{\ast}_s}^2}\,T_3(q^2)\Bigg]
\Bigg\}\,\nonumber \\
H_0^{(2)}&=& -\frac{1}{2\,M_{D^{\ast}_s}\,\sqrt{q^2}}\,(C_{10}+C_{10}^{NP}-C^{\prime}_{10})\Bigg[(M_{B_c}^2-M_{D^{\ast}_s}^2-q^2)\,
(M_{B_c}+M_{D^{\ast}_s})A_1(q^2)- \frac{\lambda}{M_{B_c}+M_{D^{\ast}_s}}\,A_2(q^2)\Bigg]\, \nonumber \\
H_t^{(1)}&=& -\sqrt{\frac{\lambda}{q^2}}\,(C_9^{\rm eff}+C_9^{NP}-C^{\prime}_9)\,A_0(q^2)\,, \nonumber \\
H_t^{(2)}&=& -\sqrt{\frac{\lambda}{q^2}}\,(C_{10}+C_{10}^{NP}-C^{\prime}_{10})\,A_0(q^2)\,,
\end{eqnarray}
where 
\begin{eqnarray}
\lambda&=& M_{B_c}^4+M_{D_s,\,D^{\ast}_s}^4+q^4-2\,(M_{B_c}^2\,M_{D_s,\,D^{\ast}_s}^2 + M_{D_s,\,D^{\ast}_s}^2\,q^2+M_{B_c}^2\,q^2)
\end{eqnarray}
Using the helicity amplitudes, the three body $B_c \to D_s\,l^+\,l^-$ and $B_c \to D^{\ast}_s\,l^+\,l^-$ differential decay rate can be
written as~\cite{Ebert:2010dv}
\begin{eqnarray}
\frac{d\Gamma}{dq^2}&=& \frac{G_F^2}{(2\pi)^3}\,\Bigg(\frac{\alpha_e\,|V_{tb}\,V_{ts}^{\ast}|}{2\pi}\Bigg)^2\,\frac{\lambda^{1/2}\,q^2}
{48\,M_{B_c}^3}\,\sqrt{1-\frac{4\,m_l^2}{q^2}}\Bigg[H^{(1)}H^{{\dagger}^{(1)}}\,\Big(1+\frac{4\,m_l^2}{q^2}\Big) + H^{(2)}H^{{\dagger}^{(2)}}\,
\Big(1-\frac{4\,m_l^2}{q^2}\Big) \, \nonumber \\
&&+ \frac{2\,m_l^2}{q^2}\,3\,H_t^{(2)}H_t^{{\dagger}^{(2)}}\Bigg]\,,
\end{eqnarray}
where $m_l$ denotes the mass of lepton and 
\begin{eqnarray}
H^{(i)}H^{{\dagger}^{(i)}} &=& H_{+}^{(i)}H_{+}^{{\dagger}^{(i)}}+H_{-}^{(i)}H_{-}^{{\dagger}^{(i)}}+H_{0}^{(i)}H_{0}^{{\dagger}^{(i)}}\,.
\end{eqnarray}
We define the differential ratio of branching ratio as follows:
\begin{eqnarray}
R_{D_s,\,D^{\ast}_s}(q^2) = \frac{d\Gamma/dq^2\,\Big(B_c \to (D_s,\,D^{\ast}_s)\mu^+\mu^-\Big)}
{d\Gamma/dq^2\,\Big(B_c \to (D_s,\,D^{\ast}_s)e^+\,e^-\Big)}\,.
\end{eqnarray}
We also construct observables like the forward backward asymmetry of the lepton pair $A_{FB}$ and the longitudinal polarization fraction
of the $D^{\ast}_s$ meson $F_L$ as a function of dilepton invariant mass $q^2$. The forward backward asymmetry $A_{FB}(q^2)$ is given by
~\cite{Ebert:2010dv}
\begin{eqnarray}
A_{FB}(q^2)&=& \frac{3}{4}\sqrt{1-\frac{4\,m_l^2}{q^2}}\Bigg\{\frac{{\rm Re}\Big(H_{+}^{(1)}H_{+}^{{\dagger}^{(2)}}\Big) - {\rm Re}
\Big(H_{-}^{(1)}H_{-}^{{\dagger}^{(2)}}\Big)}{H^{(1)}H^{{\dagger}^{(1)}}\,\Big(1+\frac{4\,m_l^2}{q^2}\Big) + H^{(2)}H^{{\dagger}^{(2)}}\,
\Big(1-\frac{4\,m_l^2}{q^2}\Big) + \frac{2\,m_l^2}{q^2}\,3\,H_t^{(2)}H_t^{{\dagger}^{(2)}}}\Bigg\}\,.
\end{eqnarray}
Similarly, the longitudinal polarization fraction of the $D^{\ast}_s$ meson can be written as~\cite{Ebert:2010dv}
\begin{eqnarray}
F_L(q^2) &=& \frac{H_0^{(1)}H_0^{{\dagger}^{(1)}}\,\Big(1+\frac{4\,m_l^2}{q^2}\Big) + H_0^{(2)}H_0^{{\dagger}^{(2)}}\,
\Big(1-\frac{4\,m_l^2}{q^2}\Big) + \frac{2\,m_l^2}{q^2}\,3\,H_t^{(2)}H_t^{{\dagger}^{(2)}}}{H^{(1)}H^{{\dagger}^{(1)}}\,\Big(1+\frac{4\,m_l^2}{q^2}\Big) + H^{(2)}H^{{\dagger}^{(2)}}\,
\Big(1-\frac{4\,m_l^2}{q^2}\Big) + \frac{2\,m_l^2}{q^2}\,3\,H_t^{(2)}H_t^{{\dagger}^{(2)}}}
\end{eqnarray}
It should be noted that the forward backward asymmetry observable $A_{FB}(q^2)$ for the $B_c \to D_s\mu^+\mu^-$ decay mode is zero in the 
SM as the helicity 
amplitudes $H_{\pm}^{i}=0$. It is worth mentioning that it can have a non zero value only if it receives contribution from scalar,
pseudoscalar or tensor NP operators. Since we consider NP in vector and axial vector operators only, we do not discuss $A_{FB}(q^2)$ for 
the $B_c \to D_s\mu^+\mu^-$ decay mode in section.~\ref{res}.

\section{Results and discussion}
\label{res}
\subsection{Inputs}
\label{in}
For definiteness, we first report all the inputs that are used for the computation of all the decay observables. We employ a
renormalization scale of $\mu = 4.8\,{\rm GeV}$ throughout our analysis. For the meson masses, we use $M_{B_c}=6.2751\,{\rm GeV}$,
$M_{D_s}=1.968\,{\rm GeV}$, and $M_{D^{\ast}_s}=2.1122\,{\rm GeV}$, as given in Ref.~\cite{Tanabashi:2018oca}. For the lepton masses, 
we use $m_e=0.5109989461\times 10^{-3}\,{\rm GeV}$ and $m_{\mu}=0.1056583715\,{\rm GeV}$ from Ref.~\cite{Tanabashi:2018oca}. 
Similarly, the mean life time of $B_c$ meson and the Fermi coupling constant are taken to be $\tau_{B_c} = 0.507\times 10^{-12}\,{\rm s}$ 
and $G_F= 1.1663787\times 10^{-5}\,{\rm GeV^{-2}}$, as reported in Ref.~\cite{Tanabashi:2018oca}. For the quark 
masses, we use $m_b(\overline{\rm MS})=4.2\,{\rm GeV}$, $m_c(\overline{\rm MS})=1.28\,{\rm GeV}$, and $m_b^{\rm pole}=4.8\,{\rm GeV}$~
\cite{Altmannshofer:2008dz}. For the electromagnetic coupling constant, 
we use $\alpha_e^{-1} = 133.28$. We use $|V_{tb}V_{ts}|=0.0401\pm 0.0010$ as given in Ref.~\cite{Bona:2006ah}. 
The WCs in our numerical estimates, taken from Refs.~\cite{Ali:1999mm}, are reported in Table.~\ref{tab_wc}.
\begin{table}[htbp]
\centering
\setlength{\tabcolsep}{8pt} 
\renewcommand{\arraystretch}{1.5} 
\begin{tabular}{|c|c|c|c|c|c|c|c|c|}
\hline
$C_1$ & $C_2$ &$ C_3$ &$ C_4$ & $C_5$ & $C_6$ &$C_7^{\rm eff}$ & $C_9$& $C_{10}$ \\
\hline
$-0.248$& $1.107$ & $0.011$ & $-0.026$ & $0.007$ & $-0.031$ & $-0.313$ & $4.344$ & $-4.669$ \\
\hline
\end{tabular}
\caption{Wilson coefficients evaluated at renormalization scale of $\mu = 4.8\,{\rm GeV}$ from Ref.~\cite{Ali:1999mm}.}
\label{tab_wc}
\end{table}
A relativistic quark model based on quasipotential approach was adopted in Ref.~\cite{Ebert:2010dv} to determine various $B_c \to D_s$ and 
$B_c \to D^{\ast}_s$ transition form factors. Various form factors at $q^2=0$ and the fitted parameters $\sigma_1$ and $\sigma_2$, taken 
from Ref.~\cite{Ebert:2010dv}, are reported in Table.~\ref{tab_ffs}.
\begin{table}[htbp]
\setlength{\tabcolsep}{8pt} 
\renewcommand{\arraystretch}{1.5} 
\begin{tabular}{|c|c|c|c|c|c|c|c|c|c|c|}
\hline
 & $f_{+}$ &$ f_0$ &$ f_T$ & $ V$ & $A_0$ &$A_1$ & $A_2$& $T_1$ & $T_2$ & $T_3$\\
\hline
$F_0$&$0.129$ &$0.129$ &$0.098$ &$0.182$ &$0.070$ &$0.089$ &$0.110$ &$0.085$ & $0.085$ & $0.051$ \\
$\sigma_1$&$2.096$ &$2.331$ &$1.412$ &$2.133$ &$1.561$ &$2.479$ &$2.833$ &$1.540$ & $2.577$ & $2.783$ \\
$\sigma_2$&$1.147$ &$1.666$ &$0.048$ &$1.183$ &$0.192$ &$1.686$ &$2.167$ &$0.248$ & $1.859$ & $2.170$ \\
\hline
\end{tabular}
\caption{$B_c \to D_s$ and $B_c \to D^{\ast}_s$ form factors at $q^2=0$ and the fitted parameters $\sigma_1$ and $\sigma_2$ from 
Ref.~\cite{Ebert:2010dv}.}
\label{tab_ffs}
\end{table}
It was shown in Ref.~\cite{Ebert:2010dv} that the $q^2$ dependence of the form factors can be well parametrized and reproduced in the form:
\begin{eqnarray}
F(q^2) &=& \frac{F(0)}{\Big(1-\frac{q^2}{M^2}\Big)\Big(1-\sigma_1\,\frac{q^2}{M_{B^{\ast}_s}^2} + \sigma_2\,\frac{q^4}{M_{B^{\ast}_s}^4}\Big)}
\end{eqnarray}
for $F(q^2) = f_{+}(q^2),\, f_T(q^2),\,V(q^2),\, A_0(q^2),\, T_1(q^2)$. Whereas, for $F(q^2) = f_{0}(q^2),\,A_1(q^2),\, A_2(q^2),\, 
T_2(q^2),\, T_3(q^2)$, it can be well approximated by
\begin{eqnarray}
F(q^2) &=& \frac{F(0)}{\Big(1-\sigma_1\,\frac{q^2}{M_{B^{\ast}_s}^2} + \sigma_2\,\frac{q^4}{M_{B^{\ast}_s}^4}\Big)}\,,
\end{eqnarray} 
where $M=M_{B_s}$ for $A_0(q^2)$ and $M=M_{B^{\ast}_s}$ for all other form factors. We use $M_{B_s} = 5.36689\,{\rm GeV}$ and
$M_{B^{\ast}_s} = 5.4154\,{\rm GeV}$ from Ref.~\cite{Tanabashi:2018oca}. The form factors describe the hadronisation of quarks 
and gluons:  these involve QCD in the non-perturbative regime and are a significant source of theoretical uncertainties. To gauge the effect
of the form factor uncertainties on various observables, we have used $\pm 5\%$ uncertainty in $F(0)$, $\sigma_1$ and $\sigma_2$.
\subsection{SM prediction of $B_c \to (D_s,\,D^{\ast}_s)\,l\,\bar l$ decay observables}
Now let us proceed to discuss our results in the SM. In Table.~\ref{tab_av_sm}, we report our $q^2$ bin averaged values of various observables
for the $B_c \to D_s\,\mu^+\,\mu^-$ and $B_c \to D^{\ast}_s\,\mu^+\,\mu^-$ decays. We restrict our analysis to low dilepton invariant
mass region and consider seven $q^2$ bins ranging from $(0.045 - 6.0)\,{\rm GeV^2}$. The central values are obtained using the central 
values of all the input parameters. For the uncertainties, we have performed a naive $\chi^2$ analysis defined as
\begin{eqnarray}
\chi^2 = \sum_{i}\,\frac{(\mathcal O_i - \mathcal O_i^0)^2}{\Delta_i^2}\,,
\end{eqnarray}
where $\mathcal O_i = (|V_{tb}V_{ts}^{\ast}|,\,F(0),\,\sigma_1,\,\sigma_2)$. Here $\mathcal O_i^0$ represents the central values of all the 
parameters and $\Delta_i$ represents $1\sigma$ uncertainty associated with each parameter. To find out the uncertainties in each observable,
We impose $\chi^2\le 2.156$ for the $B_c \to D_s\,
\mu^+\,\mu^-$ decays and $\chi^2\le 8.643$ for the $B_c \to D^{\ast}_s\,\mu^+\,\mu^-$ decays.
\begin{table}[htbp]
\centering
\setlength{\tabcolsep}{3pt} 
\renewcommand{\arraystretch}{1.5} 
\begin{tabular}{|c|c|c|c|c|c|c|c|}
\hline
Observable/$q^2$ bin & $[0.045 - 1.0]$ &$ [1.0 - 2.0]$ &$ [2.0 - 3.0]$ & $ [3.0 - 4.0]$ & $[4.0 - 5.0]$ & 
$[5.0 - 6.0]$ & $[1.0 - 6.0]$\\
\hline
\hline
$10^7 \times \mathcal B(B_c \to D_s\,\mu\mu)$ &$0.025\pm 0.001$ &$0.030\pm 0.002$ &$0.034\pm 0.002 $ &$0.038\pm 0.002$ &$0.043\pm 0.003$ &
$0.049\pm 0.004$ &$0.194\pm 0.013 $ \\
$R_{D_s}$ &$1.006\pm 0.008$ &$1.007\pm 0.002$ &$1.005\pm 0.001 $ &$1.004\pm 0.001 $ &$1.003\pm 0.001$ &$1.003\pm 0.001$ &$1.004\pm 0.001$ \\
$10^7 \times \mathcal B(B_c \to D^{\ast}_s\,\mu\mu)$ &$0.024\pm 0.001$ &$0.011\pm 0.001$ &$0.014\pm 0.002$ &$0.020\pm 0.002$ &$0.028\pm 0.003$
&$0.039\pm 0.004$ &$0.113\pm 0.012$ \\
$<A_{FB}^{D^{\ast}_s}>$ &$-0.064\pm 0.002$ &$-0.076\pm 0.012$ &$0.033\pm 0.006$ &$0.110\pm 0.009 $ &$0.160\pm 0.011 $ &$0.194\pm 0.013$ &
$0.123\pm 0.009$ \\
$<F_L^{D^{\ast}_s}>$ &$0.266\pm 0.033$ &$0.711\pm 0.034$ &$0.662\pm 0.032 $ &$0.566\pm 0.035$ &$0.488\pm 0.036 $ &$0.430\pm 0.037$ &
$0.526\pm 0.035$ \\
$R_{D^{\ast}_s}$ &$0.999\pm 0.005$ &$0.993\pm 0.002$ &$0.992\pm 0.001$ &$0.993\pm 0.001$ &$0.994\pm 0.001 $ &$0.995\pm 0.001$ &
$0.994\pm 0.001$ \\
\hline
\end{tabular}
\caption{$q^2$ bin~(in ${\rm GeV^2}$) averaged values of various observables of $B_c \to D_s\mu^+\mu^-$ and $B_c \to D^{\ast}_s\mu^+\mu^-$ 
decays in the SM. The
uncertainties in each observable corresponds to the uncertainties associated with the meson to meson transition form factors and the CKM
matrix elements.}
\label{tab_av_sm}
\end{table}
In the SM, we find the branching ratios of $B_c \to (D_s,\,D^{\ast}_s)\mu^+\mu^-$ decays to be of $\mathcal O(10^{-8})$ which might be
within the experimental sensitivity of LHCb because of the large number of $B_c$ meson that is being produced at the LHC. We also obtain
the LFU ratios to be $R_{D_s,\,D^{\ast}_s} \approx 1$ in the SM. It is observed that in the $q^2$ bin ranging from 
$(0.045 - 2)\,{\rm GeV^2}$,
the $<A_{FB}^{D^{\ast}_s}>$ observable assumes negative values, whereas, for $q^2 > 2\,{\rm GeV^2}$, it assumes positive values.
It should be noted that the uncertainty associated with the LFU ratios $R_{D_s}$ and $R_{D^{\ast}_s}$ are quite negligible in comparison
to the uncertainties present in the branching ratio, the forward backward asymmetry $<A_{FB}^{D^{\ast}_s}>$ and the longitudinal
polarization fraction of the $D^{\ast}_s$ meson $<F_L^{D^{\ast}_s}>$. Measurements of these ratios in future will be crucial in determining
various NP Lorentz structures. 

\begin{figure}[htbp]
\begin{center}
\includegraphics[width=8.9cm,height=9.0cm]{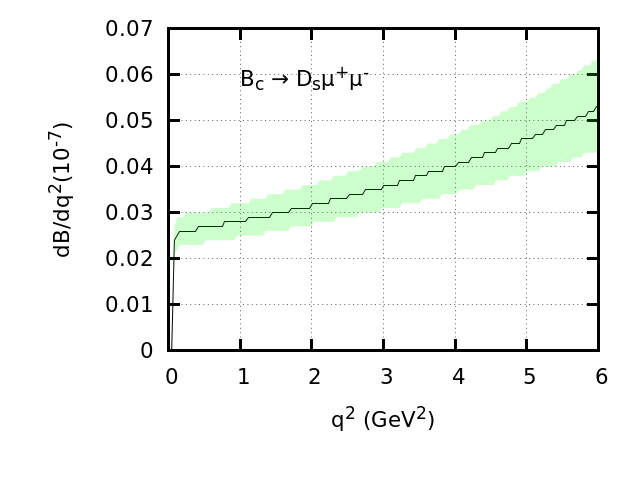}
\includegraphics[width=8.9cm,height=9.0cm]{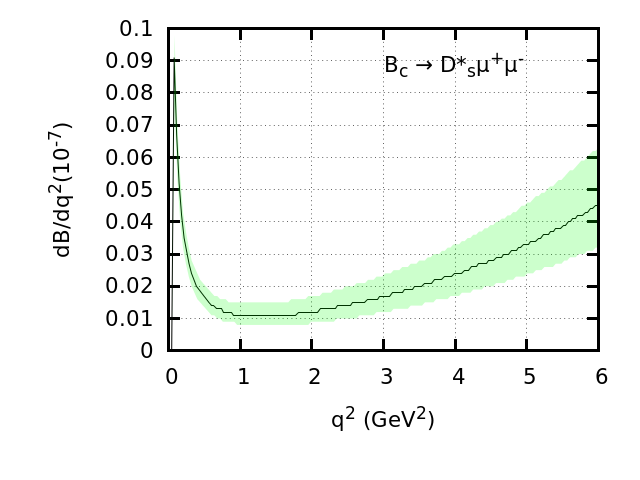}
\includegraphics[width=8.9cm,height=9.0cm]{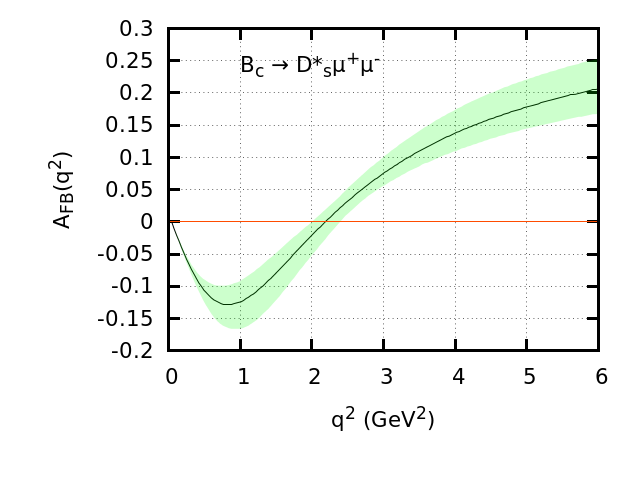}
\includegraphics[width=8.9cm,height=9.0cm]{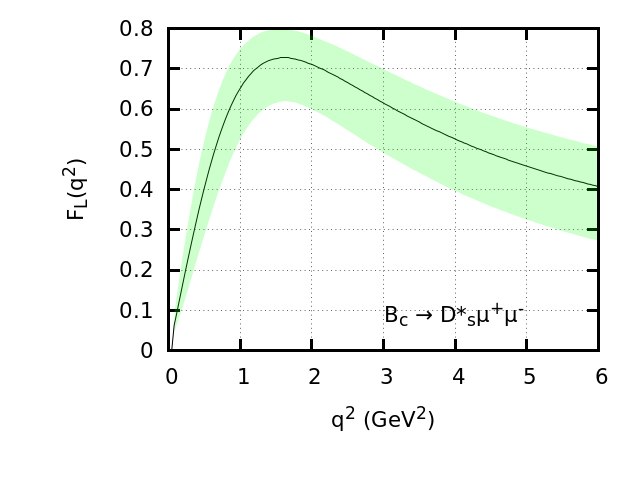}
\end{center}
\caption{Differential branching ratio~$d\mathcal B/dq^2$, forward backward asymmetry of lepton pair $A_{FB}(q^2)$ and longitudinal 
polarization 
fraction of $D^{\ast}_s$ meson $F_L(q^2)$ for the $B_c \to D_s\mu^+\mu^-$ and $B_c \to D^{\ast}_s\mu^+\mu^-$ decays in the SM. The band 
corresponds to the uncertainties in the transition form factors and the CKM matrix elements as discussed in the text.}
\label{obs_sm}
\end{figure}

We have shown in Fig.~\ref{obs_sm} the $q^2$ dependence of differential branching ratios, forward backward asymmetry, and longitudinal 
polarization fraction of $D^{\ast}_s$ meson in the low $q^2$ region $0.045 \le q^2 \le 6\,{\rm GeV^2}$. The line corresponds to the central 
values of all the input parameters, whereas, the band corresponds to the uncertainties associated with the CKM matrix element and the form 
factor inputs. In the SM, we find the zero crossing in $A_{FB}(q^2)$ of $B_c \to D^{\ast}_s\mu^+\mu^-$ decays at $q^2 = 2.2\,{\rm GeV^2}$.
Our results are quite similar to the values reported in Ref.~\cite{Ebert:2010dv}. Slight deviations may occur due to different choices of
input parameters.
\subsection{New Physics analysis}
Our main objective is to determine the effect of NP on $B_c \to (D_s,\,D^{\ast}_s)\mu^+\,\mu^-$ decay observables in a model independent way. 
To this
end, we use an effective theory formalism in the presence of new vector~(V) and axial vector~(A) couplings in our analysis. Although there
can be other NP Lorentz structures such as scalar~(S), pseudoscalar~(P) and tensor~(T), they are severely constrained by $B_s \to \mu\mu$
and $b \to s\gamma$ data. Hence we omit any discussion regarding these NP operators. Global fits of NP to the $b \to s\,l^+\,l^-$ data have 
been carried out by several groups~\cite{Capdevila:2017bsm,Altmannshofer:2017yso,
DAmico:2017mtc,Hiller:2017bzc,Geng:2017svp,Ciuchini:2017mik,Celis:2017doq,Alok:2017sui,Alok:2017jgr,Alok:2019ufo}. 
In Ref.~\cite{Alok:2019ufo}, the authors perform a global fit of 
$C_9^{NP}$, $C_{10}^{NP}$, $C^{\prime}_9$, and $C^{\prime}_{10}$ by using the
constraints coming not only from $R_K$, $R_{K^{\ast}}$, $P^{\prime}_5$, and $\mathcal B(B_s \to \phi\mu^+\mu^-)$ but also from 
$\mathcal B(B_s \to \mu^+\mu^-)$, differential branching ratios of $B^{0,\,+} \to K^{{0,\,+}^{\ast}}\mu^+\mu^-$, 
$B^{0,\,+} \to K^{0,\,+}\mu^+\mu^-$ and $B \to X_s\mu^+\mu^-$, angular observables in $B^0 \to K^{0^{\ast}}\mu^+\mu^-$ and 
$B_s \to \phi\mu^+\mu^-$ decays. Two different scenarios were considered in Ref.~\cite{Alok:2019ufo}. In $1D$ scenario, the best solutions 
to these anomalies were obtained for 
$C_9^{NP}$, $C_{10}^{NP}$, $C_9^{NP}=-C_{10}^{NP}$ and $C_9^{NP}=-C^{\prime}_9$. Similarly, for $2D$ scenario, where NP contributes to 
two WCs, the best solutions were obtained for $(C_9^{NP},\,C_{10}^{NP})$, $(C_9^{NP},\,C^{\prime}_9)$, and 
$(C_9^{NP},\,C^{\prime}_{10})$. There are other possibilities with different WCs exist that give rise to similar fits. We, however, consider
only seven of them: four from $1D$ scenario and three from $2D$ scenario. The best fit values and the corresponding $\Delta \chi^2$ values of 
all these NP WCs for $1D$ and $2D$ scenarios, taken from Ref.~\cite{Alok:2019ufo}, are reported in Table.~\ref{tab_np}. It should be noted
that NP contributions to $(C_9^{NP},\,C^{\prime}_9)$ and $(C_9^{NP},\,C^{\prime}_{10})$ are the most favored ones from the $2D$ scenario and
NP in $C_9^{NP}=-C^{\prime}_9$ is the most favored one from $1D$ scenario. 
\begin{table}[htbp]
\centering
\setlength{\tabcolsep}{8pt} 
\renewcommand{\arraystretch}{1.5} 
\begin{tabular}{|c|c|c|}
\hline
Wilson coefficients & Best fit values & $\Delta \chi^2$\\
\hline
\hline
$C_9^{NP}$ & $-1.07$ & $37.6$\\
\hline
$C_{10}^{NP}$ & $+0.78$ & $27.0$ \\
\hline
$C_9^{NP}=-C_{10}^{NP}$ & $-0.52$ & $36.3$\\
\hline
$C_9^{NP}=-C_9^{\prime}$ & $-1.11$ & $40.5$\\
\hline
\hline
$(C_9^{NP},\,C_{10}^{NP})$ & $(-0.94,\,+0.23)$ & $41.8$ \\
\hline
$(C_9^{NP},\,C_9^{\prime})$ & $(-1.27,\,+0.68)$ & $49.4$\\
\hline
$(C_9^{NP},\,C_{10}^{\prime})$ & $(-1.36,\,-0.46)$ & $52.8$ \\
\hline
\hline
\end{tabular}
\caption{Best fit and the corresponding $\Delta \chi^2$ values of different new vector and axial vector Wilson coefficients in $1D$ and 
$2D$ scenarios taken from Ref.~\cite{Alok:2019ufo}.}
\label{tab_np}
\end{table}

In Appendix~\ref{tab_app}, we report $q^2$ bin averaged values of various observables such as the branching ratio, ratio of branching
ratio, forward backward asymmetry, and longitudinal polarization fraction of the $D^{\ast}_s$ meson for the $B_c \to D_s\mu^+\mu^-$ 
and $B_c \to D^{\ast}_s\mu^+\mu^-$ decays in the presence of these NP WCs. In each $q^2$ bin, the branching ratio of 
$B_c \to D_s\mu^+\mu^-$ is smaller
in each NP scenarios than in the SM except for $C_9^{NP}=-C^{\prime}_9$. It remains SM like for $C_9^{NP}=-C^{\prime}_9$. Similar
conclusion can be made for the ratio of branching ratio $R_{D_s}$ as well. For the $B_c \to D^{\ast}_s\mu^+\mu^-$ decay, the bin averaged
branching ratio and the ratio of branching ratio $R_{D^{\ast}_s}$ in each $q^2$ bin for each NP scenarios are smaller than the 
corresponding SM value. However, the $A_{FB}$ and $F_L$ values can be either smaller or larger in NP cases than the SM central 
value. It should be mentioned that the deviation of $R_{D_s}$, $R_{D^{\ast}_s}$ and $A_{FB}$ from the SM prediction can be quite large in 
some $q^2$ bins.

\begin{figure}[htbp]
\begin{center}
\includegraphics[width=8.9cm,height=9.0cm]{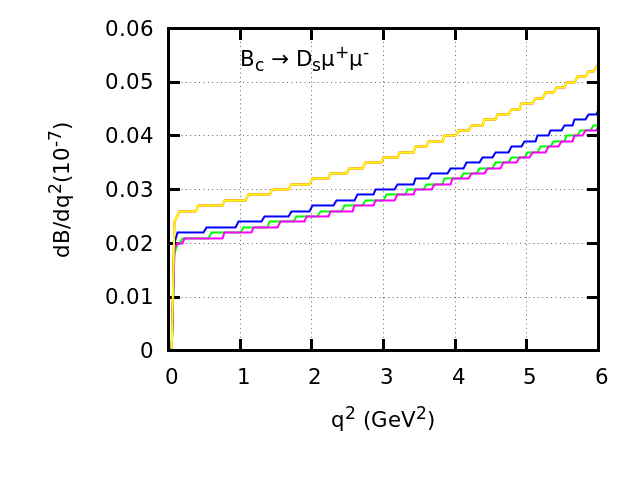}
\includegraphics[width=8.9cm,height=9.0cm]{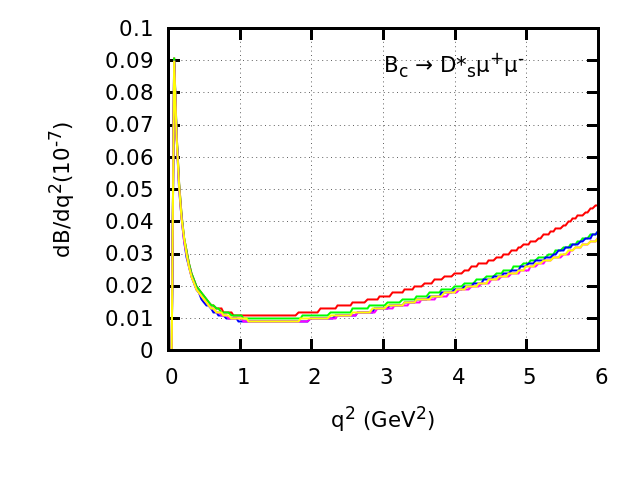}
\includegraphics[width=8.9cm,height=9.0cm]{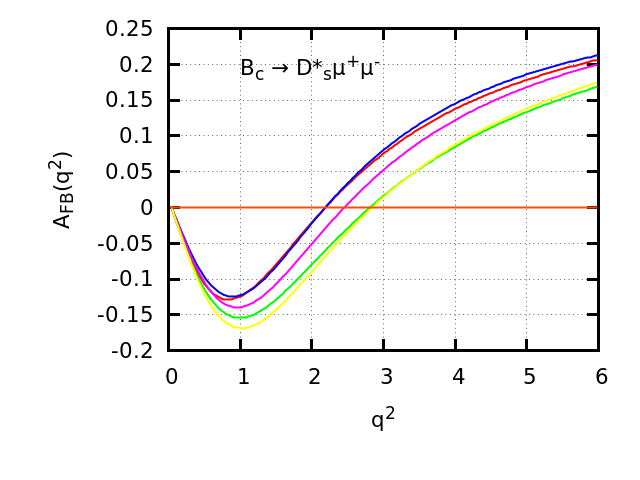}
\includegraphics[width=8.9cm,height=9.0cm]{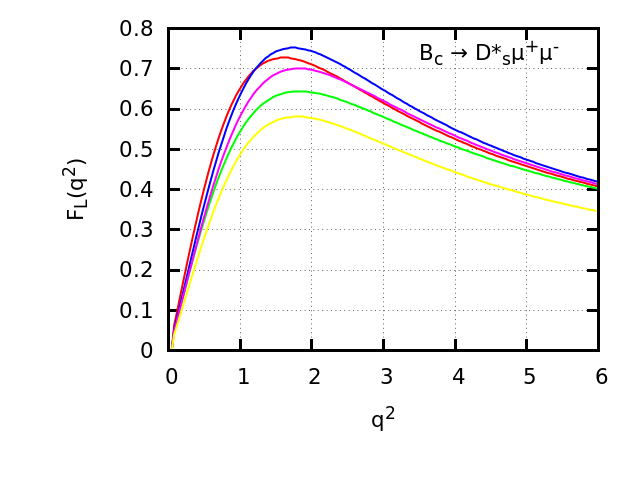}
\end{center}
\caption{Differential branching ratio~$d\mathcal B/dq^2$, forward backward asymmetry of lepton pair $A_{FB}(q^2)$ and longitudinal 
polarization 
fraction of $D^{\ast}_s$ meson $F_L(q^2)$ for the $B_c \to D_s\mu^+\mu^-$ and $B_c \to D^{\ast}_s\mu^+\mu^-$ decays in the SM~(Red) and
for the best fit values of new VA couplings in $1D$ scenario. Green, blue, purple, and yellow lines correspond to the best fit values of
$C_9^{NP}=-1.07$, $C_{10}^{NP}=0.78$, $C_9^{NP}=-C_{10}^{NP}=-0.52$, and $C_9^{NP}=-C^{\prime}_9=-1.11$, respectively.}
\label{obs_1d}
\end{figure}

In Fig.~\ref{obs_1d}, we show various $q^2$ dependent observables for the $B_c \to (D_s,\,D^{\ast}_s)\mu^+\,\mu^-$ decays in the presence
of various NP WCs in $1D$ scenario. Our observations are as follows:
\begin{itemize}
\item
The differential branching ratio for the $B_c \to D_s\,\mu^+\,\mu^-$ decays
is reduced at all $q^2$ for $C_9^{NP}$, $C_{10}^{NP}$, and $C_9^{NP}=-C_{10}^{NP}$, whereas, it remains SM like for
$C_9^{NP}=-C^{\prime}_9$. This could very well be understood from Eq.~\ref{dseq} that $H_0^{(1)}$ and $H_t^{(1)}$ helicity amplitudes
for the $B_c \to D_s\,\mu^+\,\mu^-$ decay mode depend on the combination $C_9^{NP}+C^{\prime}_9$. Hence the NP contribution cancels.

\item
The differential branching ratio for the $B_c \to D^{\ast}_s\,\mu^+\,\mu^-$ decays
is reduced at all $q^2$ for $C_9^{NP}$, $C_{10}^{NP}$, $C_9^{NP}=-C_{10}^{NP}$, and $C_9^{NP}=-C^{\prime}_9$. The deviation with the SM
prediction increases as $q^2$ increases for each NP WCs.

\item
For all the NP couplings, the zero crossing in the forward backward asymmetry observable $A_{FB}^{D^{\ast}_s}$ is shifted to the higher values
of $q^2$ than in the SM. There is, however, one exception. For $C_{10}^{NP}=0.78$, the zero crossing coincides with the SM prediction 
although the shape of 
$A_{FB}^{D^{\ast}_s}$ may slightly vary. Maximum deviation from the SM prediction is observed for $C_9^{NP}$ and $C_9^{NP}=-C^{\prime}_9$.

\item
The peak of the longitudinal polarization fraction of $D^{\ast}_s$ meson may shift towards a higher values of $q^2$ than in the SM. Although
the longitudinal polarization fraction $F_L$ is reduced at all $q^2$ for $C_9^{NP}$, $C_9^{NP}=-C_{10}^{NP}$, and $C_9^{NP}=-C^{\prime}_9$,
it may increase with $C_{10}^{NP}$ for $q^2 > 1.2\,{\rm GeV^2}$.
\end{itemize}
There are other combinations of VA couplings exist in the $1D$ scenario as reported in Ref.~\cite{Alok:2019ufo}. We, however, do not 
consider those cases because of their small $\Delta \chi^2$ values. 

We now consider several NP couplings from the $2D$ scenarios having high $\Delta \chi^2$ values from the global fit~\cite{Alok:2019ufo}. 
The best fit values, taken from Ref.~\cite{Alok:2019ufo}, are reported in table.~\ref{tab_np}. We show in Fig.~\ref{obs_2d} various 
observables such as Differential branching ratio~$d\mathcal B/dq^2$, forward backward asymmetry of lepton pair $A_{FB}(q^2)$ and 
longitudinal 
polarization fraction of $D^{\ast}_s$ meson $F_L(q^2)$ as a function of dilepton invariant mass $q^2$ for the $B_c \to D_s\mu^+\mu^-$ and 
$B_c \to D^{\ast}_s\mu^+\mu^-$ decays in the presence of such NP. Our main observations are as follows:
\begin{figure}[htbp]
\begin{center}
\includegraphics[width=8.9cm,height=9.0cm]{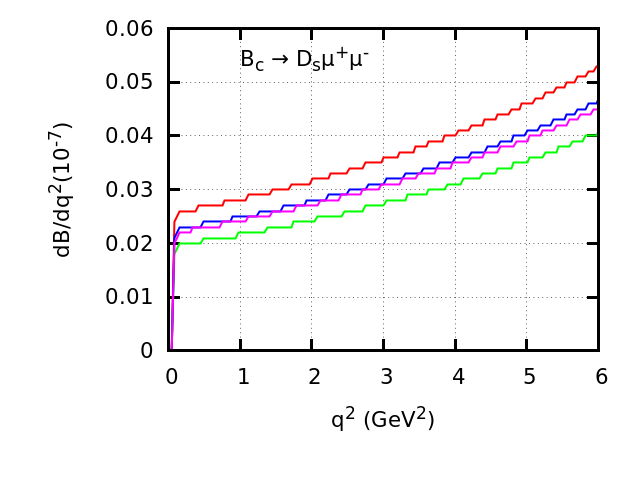}
\includegraphics[width=8.9cm,height=9.0cm]{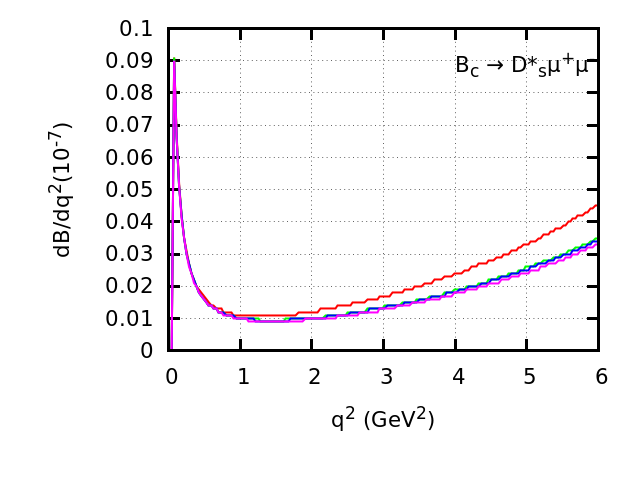}
\includegraphics[width=8.9cm,height=9.0cm]{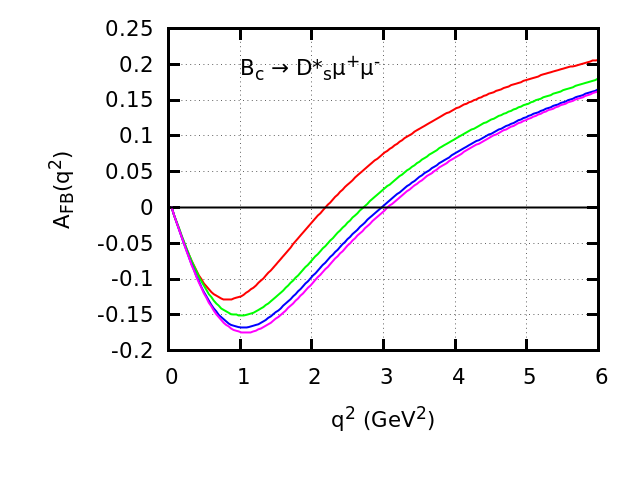}
\includegraphics[width=8.9cm,height=9.0cm]{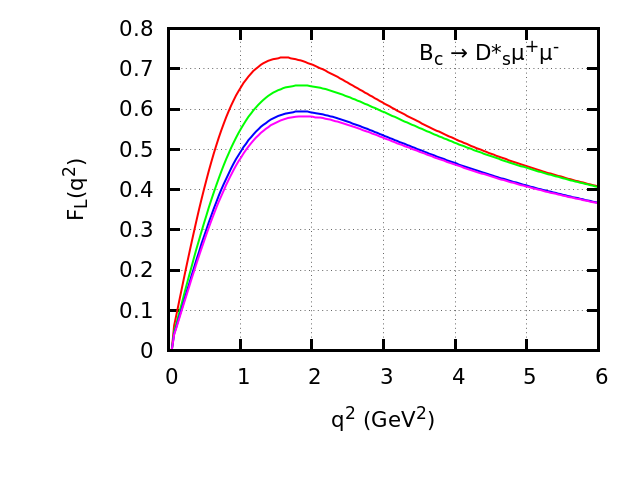}
\end{center}
\caption{Differential branching ratio~$d\mathcal B/dq^2$, forward backward asymmetry of lepton pair $A_{FB}(q^2)$ and longitudinal 
polarization
fraction of $D^{\ast}_s$ meson $F_L(q^2)$ for the $B_c \to D_s\mu^+\mu^-$ and $B_c \to D^{\ast}_s\mu^+\mu^-$ decays in the SM~(Red) and
for the best fit values of new VA couplings in $2D$ scenario. Green, blue and purple lines correspond to the best fit values of
$(C_9^{NP},\,C_{10}^{NP})=(-0.94,\,+0.23)$, $(C_9^{NP},\,C^{\prime}_9)=(-1.27,\,+0.68)$ and $(C_9^{NP},\,C^{\prime}_{10})=(-1.36,\,-0.46)$, 
respectively.}
\label{obs_2d}
\end{figure}
\begin{itemize}
\item
The differential branching ratio for the $B_c \to D_s\mu^+\mu^-$ decay is reduced at all $q^2$ for each NP couplings. The deviation from
the SM prediction is more pronounced in case of $(C_9^{NP},\,C_{10}^{NP})=(-0.94,\,+0.23)$.

\item
Similar to $B_c \to D_s\mu^+\mu^-$, the differential branching ratio for the $B_c \to D^{\ast}_s\mu^+\mu^-$ decay also is reduced at all
$q^2$. The deviation with the SM prediction, however, increases with increase in $q^2$. It reaches maximum at $q^2 = 6\,{\rm GeV^2}$.

\item
The zero crossing in the forward backward asymmetry observable $A_{FB}(q^2)$ is shifted to higher values of $q^2$ than in the SM for each
NP couplings. The maximum deviation from the SM prediction is observed for $(C_9^{NP},\,C^{\prime}_{10})=(-1.36,\,-0.46)$ which is shown with 
a purple line in Fig.~\ref{obs_2d}.

\item
The longitudinal polarization fraction of the $D^{\ast}_s$ meson $F_L(q^2)$ decreases once we include the NP couplings. It is observed that
the peak of the $F_L(q^2)$ distribution reduces and shifted towards slightly higher $q^2$ than in the SM. Maximum deviation from the SM
prediction is observed for $(C_9^{NP},\,C^{\prime}_{10})=(-1.36,\,-0.46)$ which is shown with a purple line in Fig.~\ref{obs_2d}.
\end{itemize}

\section{conclusion}
\label{con}
Motivated by the anomalies present in $B \to (K,\,K^{\ast})\mu^+\mu^-$ decays,  
we have analyzed $B_c \to (D_s,\,D^{\ast}_s)\mu^+\mu^-$ decays mediated via $b \to s\,l^+\,l^-$ neutral current transitions using the 
$B_c \to (D_s,\,D^{\ast}_s)$ transition form factors obtained in the relativistic quark model. We use a model independent effective 
theory formalism and include NP effects coming from new vector and axial vector operators only.
We discard scalar, pseudoscalar and tensor operators as they are severely constrained by $B_s \to \mu\mu$ and $b \to s\gamma$ data. Several
authors have performed global fits to the $b \to s\,l^+\,l^-$ data and proposed two types of NP scenarios, namely, $1D$ and $2D$ scenarios. 
In the $1D$ scenario, we chose four NP scenarios
in which the NP contribution is coming from only one NP WCs at a time. In the $2D$ scenario, we chose three NP scenarios in which the
NP contribution is coming from two NP WCs at a time.

We give predictions on several observables such as branching ratio, ratio of branching ratio, forward backward asymmetry, and the longitudinal
polarization fraction of the $D^{\ast}_s$ meson in the SM and in several NP cases. We observe that for most of the NP cases, the branching
ratio for both the decay modes is reduced at all $q^2$. In most cases, the zero of $A_{FB}(q^2)$ parameter is shifted to the higher value of
$q^2$ than in the SM. However, with only $C_{10}^{NP}$, the zero crossing is SM like. Similarly, for the longitudinal polarization fraction
of the $D^{\ast}_s$ meson, the peak of the distribution is reduced at all $q^2$ and it is slightly shifted towards higher value of $q^2$ than 
in the SM.

Although there is hint of NP in $b \to s\,l^+\,l^-$ transition decays, NP is not yet established. 
Unlike $B \to (K,\,K^{\ast})\mu^+\mu^-$ decays which are rigorously studied both theoretically and experimentally, the 
$B_c \to (D_s,\,D^{\ast}_s)\mu^+\mu^-$ decays mediated via same $b \to s\,l^+\,l^-$ neutral current transitions received very less attention.
Measurement of various observables for the $B_c \to (D_s,\,D^{\ast}_s)\mu^+\mu^-$ decays and at the same time improved estimates of various 
$B_c \to D_s$ and $B_c \to D^{\ast}_s$ transition form factors in future will be crucial in identifying the true nature of NP. Again,
to enhance the significance of various measurements related to $b \to s\,l^+\,l^-$ decays and to disentangle genuine NP effects from various
statistical and systematic uncertainties, more data samples are needed.

\appendix*
\label{app}
\section{}
\label{tab_app}
Here we report the $q^2$ bin averaged values of all the observables for the $B_c \to (D_s,\,D^{\ast}_s)\mu^+\mu^-$ decays in the SM and 
in several NP cases.
\begin{table}[htbp]
\centering
\setlength{\tabcolsep}{8pt} 
\renewcommand{\arraystretch}{1.5} 
\begin{tabular}{|c|c|c|c|c|c||c|c|c|}
\hline
$q^2$ bin~(${\rm GeV^2})$ & SM &$C_9^{NP} $ &$ C_{10}^{NP}$ & $ C_9^{NP}=-C_{10}^{NP}$ & $C_9^{NP}=-C^{\prime}_9$ &$(C_9^{NP},\,C_{10}^{NP})$ &
$(C_9^{NP},\,C^{\prime}_9)$ & $(C_9^{NP},\,C^{\prime}_{10})$\\
\hline
\hline
$[0.045,1.0]$ & $0.025$ &$0.020$ &$0.021$ &$0.020$ & $0.025$ &$0.019$ &$0.022$ & $0.022$\\
$[1.0,2.0]$ & $0.030$ & $0.024$ & $0.025$ & $0.023$ & $0.030$ &$0.023$ &$0.026$ & $0.026$\\
$[2.0,3.0]$ & $0.034$ & $0.027$ & $0.028$ & $0.026$ & $0.034$ &$0.026$ &$0.030$ &$0.029$\\
$[3.0,4.0]$ & $0.038$ & $0.030$ & $0.032$ & $0.030$ & $0.038$ &$0.029$ &$0.033$ &$0.033$\\
$[4.0,5.0]$ & $0.043$ & $0.034$ & $0.036$ & $0.034$ & $0.043$ &$0.033$ &$0.038$ &$0.037$\\
$[5.0,6.0]$ & $0.049$ & $0.039$ & $0.042$ & $0.039$ & $0.049$ &$0.038$ &$0.043$ &$0.042$\\
$[1.0,6.0]$ & $0.193$ & $0.154$ & $0.163$ & $0.152$ & $0.193$ &$0.149$ &$0.171$ &$0.166$\\
\hline
\hline
\end{tabular}
\caption{$q^2$ bin averaged values of $10^7\times\mathcal B(B_c \to D_s\mu^+\mu^-)$ in the SM and in several NP cases from $1D$ and $2D$ 
scenarios.}
\label{ds_br_av_np}
\end{table}
\begin{table}[htbp]
\centering
\setlength{\tabcolsep}{8pt} 
\renewcommand{\arraystretch}{1.5} 
\begin{tabular}{|c|c|c|c|c|c||c|c|c|}
\hline
$q^2$ bin~(${\rm GeV^2})$ & SM &$C_9^{NP} $ &$ C_{10}^{NP}$ & $ C_9^{NP}=-C_{10}^{NP}$ & $C_9^{NP}=-C^{\prime}_9$ &$(C_9^{NP},\,C_{10}^{NP})$ &
$(C_9^{NP},\,C^{\prime}_9)$ & $(C_9^{NP},\,C^{\prime}_{10})$\\
\hline
\hline
$[0.045,1.0]$ & $1.006$ &$0.798$ &$0.846$ &$0.788$ & $1.006$ & $0.770$ &$0.884$ &$0.859$\\
$[1.0,2.0]$ & $1.007$ & $0.803$ & $0.845$ & $0.790$ & $1.007$ &$0.774$ &$0.888$ &$0.867$ \\
$[2.0,3.0]$ & $1.005$ & $0.802$ & $0.844$ & $0.788$ & $1.005$&$0.773$ &$0.886$& $0.865$ \\
$[3.0,4.0]$ & $1.004$ & $0.801$ & $0.844$ & $0.788$ & $1.004$ &$0.772$ &$0.885$ &$0.863$ \\
$[4.0,5.0]$ & $1.003$ & $0.800$ & $0.845$ & $0.789$ & $1.003$ &$0.772$ &$0.884$ &$0.861$\\
$[5.0,6.0]$ & $1.003$ & $0.799$ & $0.847$ & $0.790$ & $1.003$ &$0.772$ &$0.884$ &$0.859$\\
$[1.0,6.0]$ & $1.004$ & $0.801$ & $0.845$ & $0.789$ & $1.004$ &$0.773$ &$0.885$ &$0.863$\\
\hline
\hline
\end{tabular}
\caption{$q^2$ bin averaged values of $R_{D_s}$ in the SM and in several NP cases from $1D$ and $2D$ scenarios.}
\label{ds_r_av_np}
\end{table}
\begin{table}[htbp]
\centering
\setlength{\tabcolsep}{8pt} 
\renewcommand{\arraystretch}{1.5} 
\begin{tabular}{|c|c|c|c|c|c||c|c|c|}
\hline
$q^2$ bin~(${\rm GeV^2})$ & SM &$C_9^{NP} $ &$ C_{10}^{NP}$ & $ C_9^{NP}=-C_{10}^{NP}$ & $C_9^{NP}=-C^{\prime}_9$ &$(C_9^{NP},\,C_{10}^{NP})$ &
$(C_9^{NP},\,C^{\prime}_9)$ & $(C_9^{NP},\,C^{\prime}_{10})$\\
\hline
\hline
$[0.045,1.0]$ & $0.024$ &$0.024$ &$0.023$ &$0.023$ & $0.023$ &$0.023$ &$0.023$ & $0.023$\\
$[1.0,2.0]$ & $0.011$ & $0.010$ & $0.009$ & $0.009$ & $0.009$ &$0.010$ &$0.010$ & $0.009$\\
$[2.0,3.0]$ & $0.014$ & $0.012$ & $0.011$ & $0.011$ & $0.011$ &$0.012$ &$0.011$ &$0.011$\\
$[3.0,4.0]$ & $0.020$ & $0.017$ & $0.016$ & $0.015$ & $0.016$ &$0.016$ &$0.016$ &$0.015$\\
$[4.0,5.0]$ & $0.028$ & $0.023$ & $0.022$ & $0.022$ & $0.022$ &$0.022$ &$0.022$ &$0.021$\\
$[5.0,6.0]$ & $0.039$ & $0.032$ & $0.031$ & $0.030$ & $0.030$ &$0.030$ &$0.030$ &$0.028$\\
$[1.0,6.0]$ & $0.113$ & $0.095$ & $0.090$ & $0.087$ & $0.088$ &$0.089$ &$0.088$ &$0.085$\\
\hline
\hline
\end{tabular}
\caption{$q^2$ bin averaged values of $10^7\times\mathcal B(B_c \to D^{\ast}_s\mu^+\mu^-)$ in the SM and in several NP cases from $1D$ and 
$2D$ scenarios.}
\label{dst_br_av_np}
\end{table}
\begin{table}[htbp]
\centering
\setlength{\tabcolsep}{8pt} 
\renewcommand{\arraystretch}{1.5} 
\begin{tabular}{|c|c|c|c|c|c||c|c|c|}
\hline
$q^2$ bin~(${\rm GeV^2})$ & SM &$C_9^{NP} $ &$ C_{10}^{NP}$ & $ C_9^{NP}=-C_{10}^{NP}$ & $C_9^{NP}=-C^{\prime}_9$ &$(C_9^{NP},\,C_{10}^{NP})$ &
$(C_9^{NP},\,C^{\prime}_9)$ & $(C_9^{NP},\,C^{\prime}_{10})$\\
\hline
\hline
$[0.045,1.0]$ & $0.999$ &$0.995$ &$0.994$ &$0.957$ & $0.961$ &$0.977$ &$0.975$ & $0.970$\\
$[1.0,2.0]$ & $0.993$ & $0.915$ & $0.801$ & $0.815$ & $0.834$ &$0.860$ &$0.855$ & $0.830$\\
$[2.0,3.0]$ & $0.992$ & $0.859$ & $0.776$ & $0.770$ & $0.790$ &$0.804$ &$0.797$ &$0.767$\\
$[3.0,4.0]$ & $0.993$ & $0.834$ & $0.781$ & $0.760$ & $0.778$ &$0.783$ &$0.774$ &$0.744$\\
$[4.0,5.0]$ & $0.994$ & $0.820$ & $0.790$ & $0.760$ & $0.772$ &$0.774$ &$0.763$ &$0.733$\\
$[5.0,6.0]$ & $0.995$ & $0.812$ & $0.799$ & $0.763$ & $0.768$ &$0.769$ &$0.756$ &$0.726$\\
$[1.0,6.0]$ & $0.994$ & $0.834$ & $0.791$ & $0.768$ & $0.786$ &$0.786$ &$0.776$ &$0.746$\\
\hline
\hline
\end{tabular}
\caption{$q^2$ bin averaged values of $R_{D^{\ast}_s}$ in the SM and in several NP cases from $1D$ and $2D$ scenarios.}
\label{dst_r_av_np}
\end{table}
\begin{table}[htbp]
\centering
\setlength{\tabcolsep}{8pt} 
\renewcommand{\arraystretch}{1.5} 
\begin{tabular}{|c|c|c|c|c|c||c|c|c|}
\hline
$q^2$ bin~(${\rm GeV^2})$ & SM &$C_9^{NP} $ &$ C_{10}^{NP}$ & $ C_9^{NP}=-C_{10}^{NP}$ & $C_9^{NP}=-C^{\prime}_9$ &$(C_9^{NP},\,C_{10}^{NP})$ &
$(C_9^{NP},\,C^{\prime}_9)$ & $(C_9^{NP},\,C^{\prime}_{10})$\\
\hline
\hline
$[0.045,1.0]$ & $-0.064$ &$-0.069$ & $-0.056$ &$-0.062$ &$-0.072$ & $-0.066$ &$-0.072$ &$-0.072$ \\
$[1.0,2.0]$ & $-0.076$ & $-0.124$ & $-0.079$ & $-0.103$ & $-0.138$ &$-0.121$ &$-0.141$ & $-0.150$\\
$[2.0,3.0]$ & $0.033$ & $-0.028$ & $0.035$ & $0.006$ & $-0.033$ &$-0.020$ &$-0.043$ &$-0.052$\\
$[3.0,4.0]$ & $0.110$ & $0.055$ & $0.117$ & $0.092$ & $0.055$ &$0.065$ &$0.043$ &$0.037$\\
$[4.0,5.0]$ & $0.160$ & $0.112$ & $0.168$ & $0.148$ & $0.116$ &$0.123$ &$0.104$ &$0.099$\\
$[5.0,6.0]$ & $0.194$ & $0.153$ & $0.201$ & $0.186$ & $0.159$ &$0.164$ &$0.147$ &$0.144$\\
$[1.0,6.0]$ & $0.123$ & $0.072$ & $0.129$ & $0.107$ & $0.074$ &$0.082$ &$0.062$ &$0.056$\\
\hline
\hline
\end{tabular}
\caption{$q^2$ bin averaged values of $<A_{FB}^{D^{\ast}_s}>$ in the SM and in several NP cases from $1D$ and $2D$ scenarios.}
\label{dst_afb_av_np}
\end{table}
\begin{table}[htbp]
\centering
\setlength{\tabcolsep}{8pt} 
\renewcommand{\arraystretch}{1.5} 
\begin{tabular}{|c|c|c|c|c|c||c|c|c|}
\hline
$q^2$ bin~(${\rm GeV^2})$ & SM &$C_9^{NP} $ &$ C_{10}^{NP}$ & $ C_9^{NP}=-C_{10}^{NP}$ & $C_9^{NP}=-C^{\prime}_9$ &$(C_9^{NP},\,C_{10}^{NP})$ &
$(C_9^{NP},\,C^{\prime}_9)$ & $(C_9^{NP},\,C^{\prime}_{10})$\\
\hline
\hline
$[0.045,1.0]$ & $0.266$ &$0.213$ &$0.232$ &$0.215$ & $0.181$ &$0.208$ &$0.186$ & $0.177$\\
$[1.0,2.0]$ & $0.711$ & $0.620$ & $0.725$ & $0.671$ & $0.558$ &$0.630$ &$0.567$ & $0.554$\\
$[2.0,3.0]$ & $0.662$ & $0.613$ & $0.697$ & $0.661$ & $0.548$ &$0.627$ &$0.566$ &$0.558$\\
$[3.0,4.0]$ & $0.566$ & $0.541$ & $0.594$ & $0.573$ & $0.476$ &$0.552$ &$0.498$ &$0.493$\\
$[4.0,5.0]$ & $0.488$ & $0.475$ & $0.508$ & $0.496$ & $0.413$ &$0.482$ &$0.435$ &$0.432$\\
$[5.0,6.0]$ & $0.430$ & $0.423$ & $0.444$ & $0.437$ & $0.365$ &$0.428$ &$0.387$ &$0.385$\\
$[1.0,6.0]$ & $0.526$ & $0.503$ & $0.546$ & $0.529$ & $0.440$ &$0.511$ &$0.461$ &$0.457$\\
\hline
\hline
\end{tabular}
\caption{$q^2$ bin averaged values of $<F_L^{D^{\ast}_s}>$ in the SM and in several NP cases from $1D$ and $2D$ scenarios.}
\label{dst_fl_av_np}
\end{table}
%

\end{document}